\documentclass[prb,twocolumn,amsmath,amssymb,superscriptaddress]{revtex4}

\usepackage{graphicx} \ifx\pdfoutput\undefined 
\usepackage[ps2pdf]{thumbpdf} \DeclareGraphicsExtensions{.eps} \else \DeclareGraphicsExtensions{.png,.pdf} 
\usepackage{epstopdf} 
\usepackage[pdftex]{thumbpdf} \fi

\usepackage{epsfig,psfrag,amsmath,amssymb,float}

\begin{document}

\newcommand{\ltwid}{\mathrel{\raise.3ex\hbox{$<$\kern-.75em\lower1ex\hbox{$\sim$}}}} 
\newcommand{\gtwid}{\mathrel{\raise.3ex\hbox{$>$\kern-.75em\lower1ex\hbox{$\sim$}}}} 
\def\K{{\bf{K}}} 
\def\Q{{\bf{Q}}} 
\def\Gbar{\bar{G}} 
\def\tk{\tilde{\bf{k}}} 
\def\k{{\bf{k}}} 
\def\kt{{\tilde{\bf{k}}}} 
\def\p{{\bf{p}}} 
\def\q{{\bf{q}}} 
\def\pp{{\bf{p}}^\prime} 
\def\Gpp{\Gamma^{pp}} 
\def\Phid{\Phi_d(\k,\omega_n)} 
\def\ld{\lambda_d(T)} 
\def\n{\langle n \rangle } 
\def\dw{d_{x^2-y^2}} 
\def\Ub{\bar{U}}

\title{Systematic analysis of a spin-susceptibility representation of the pairing interaction in the 2D Hubbard model}

\author{T.A.~Maier} \email{maierta@ornl.gov} \affiliation{Computer Science and Mathematics Division, Oak Ridge National Laboratory, Oak Ridge, TN 37831-6164} \affiliation{Center for Nanophase Materials Sciences, Oak Ridge National Laboratory, Oak Ridge, TN 37831-6164} 

\author{A. Macridin} \email{macridin@physics.uc.edu} \affiliation{Department of Physics, University of Cincinnati, Cincinnati, OH 45221} 

\author{M.~Jarrell} \email{jarrell@physics.uc.edu} \affiliation{Department of Physics, University of Cincinnati, Cincinnati, OH 45221} 

\author{D.J.~Scalapino} \email{djs@.physics.ucsb.edu} \affiliation{Department of Physics, University of California, Santa Barbara, CA 93106-9530} \affiliation{Center for Nanophase Materials Sciences, Oak Ridge National Laboratory, Oak Ridge, TN 37831-6164}  

\date{\today} 
\begin{abstract}

	A dynamic cluster quantum Monte Carlo algorithm is used to study a spin susceptibility representation of the pairing interaction for the two-dimensional Hubbard model with an on-site Coulomb interaction equal to the bandwidth for various doping levels. We find that the pairing interaction is well approximated by $\frac{3}{2}\Ub(T)^2\chi(K-K')$ with an effective temperature and doping dependent coupling $\Ub(T)$ and the numerically calculated spin susceptibility $\chi(K-K')$. We show that at low temperatures, $\Ub$ may be accurately determined from a corresponding spin susceptibility based calculation of the single-particle self-energy. We conclude that the strength of the d-wave pairing interaction, characterized by the mean-field transition temperature, can be determined from a knowledge of the dressed spin susceptibility and the nodal quasiparticle spectral weight. This has important implications with respect to the questions of whether spin fluctuations are responsible for pairing in the high-T$_c$ cuprates.

\end{abstract}

\maketitle

\section{Introduction}

Recent numerical calculations have shown that the dominant contribution to the d-wave pairing interaction in the 2D Hubbard model comes from the spin $S=1$ channel \cite{maier:pairmech,maier:pairint}. Motivated by this result, a simple spin susceptibility representation of the pairing interaction was studied \cite{maier:07}. Results for a Hubbard on-site Coulomb interaction equal to the bandwidth and a site filling $\n=0.85$ have shown that the pairing interaction can be well approximated by a simple RPA form \cite{berk:66,scalapino:95,moriya:00,chubokov:03}
\begin{equation} \label{eq:ppRPA} 
	\frac{3}{2} \Ub^2(T) \chi(K-K')\,. 
\end{equation}
Here it was important that an effective temperature dependent coupling $\Ub(T)$ and the dressed spin susceptibility were used in Eq.~\eqref{eq:ppRPA} instead of the bare $U$ and the perturbative RPA susceptibility. The coupling $\Ub(T)$ was determined by fitting the low frequency d-wave projected irreducible particle-particle vertex calculated with a dynamic cluster approximation (DCA) quantum Monte Carlo (QMC) technique \cite{hettler:dca1,hettler:dca2,jarrell:dca3,maier:rev} with the d-wave projection of the form given by Eq.~\eqref{eq:ppRPA}. Using this estimate of $\Ub(T)$ and the calculated dressed susceptibility and Green's function, it was shown that the eigenvalue and eigenfunction of the homogeneous Bethe-Salpeter equation in the particle-particle channel are well represented by the corresponding quantities calculated with the approximate interaction given by Eq.~\eqref{eq:ppRPA}.

Here we extend this study to explore this approximation for other fillings and investigate other ways to determine the coupling strength $\Ub(T)$ that do not require knowledge of the irreducible particle-particle vertex, since this is not experimentally accessible for the cuprates. We first discuss the DCA QMC technique used to calculate the relevant quantities and review the fitting procedure used to determine the temperature dependent coupling $\Ub(T)$ from the irreducible particle-particle vertex. We then examine how well Eq.~\eqref{eq:ppRPA} can describe the pairing interaction for various site fillings $\n$. We are particularly interested in the low doping regime. In this case, a pseudogap opens in the density of states and the low-energy spin excitations \cite{maier:dca1,jarrell:dca2,macridin:06,senechal:04a,kyung:06} and it is unclear whether the form given by Eq.~\eqref{eq:ppRPA} is still a good representation of the pairing interaction. We then explore two approximations that estimate $\Ub(T)$ from the single-particle spectrum. By assuming that the self-energy is determined by the same interaction [Eq.~\eqref{eq:ppRPA}] as the particle-particle interaction, one can get a single-particle estimate for $\Ub(T)$. We first do this assuming that the exact single-particle spectral weight is known. Next we consider a scenario, where only limited information is available for the single-particle spectrum, such as the nodal spectral weight. In this case, one can use the interaction in Eq.~\eqref{eq:ppRPA} to self-consistently determine the self-energy and dressed Green's function in addition to the effective coupling $\Ub(T)$. This, of course, assumes that a reliable estimate of the non-interacting Green's function is available. 

We will study the quality of these approximations by comparing their respective estimates of the d-wave eigenvalue with the ``exact'' result. Throughout this paper, ``exact'' will refer to the numerical results obtained using the DCA QMC technique. These DCA QMC calculations are carried out on a 4-site cluster. This means that phase fluctuations are suppressed and the temperature $T_{c0}$ at which the d-wave eigenvalue equals one corresponds to the mean-field transition temperature \cite{maier:schm}. This temperature provides a natural measure of the strength of the d-wave pairing interaction.

\section{Dynamic Cluster quantum Monte Carlo technique}\label{sec:DCA}

To calculate the single-particle self-energy, the pairing interaction and the spin susceptibility in the 2D Hubbard model, we use a DCA QMC algorithm \cite{hettler:dca1,hettler:dca2,jarrell:dca3,maier:rev}. The dynamic cluster approximation maps the original lattice model onto a periodic cluster of size $N_c$ sites embedded in a self-consistent host. The essential assumption is that short-range quantities, such as the self energy and its functional derivatives (the irreducible vertex functions) are well represented as diagrams constructed from the coarse-grained Green's function. For the problem of interest, this is a reasonable assumption for systems where the correlations that mediate the pairing are short-ranged. To this end, the first Brillouin zone is divided into $N_c$ cells, with each cell represented by its center wave-vector $\K$ surrounded by $N/N_c$ lattice wavevectors labeled by $\tk$. The reduction of the $N$-site lattice problem to an effective $N_c$ site cluster problem is achieved by coarse-graining the single-particle Green's function, {\it i.e.} averaging $G(\K+\tilde{\k})$ over the $\tk$ within a cell which converges to a cluster Green's function $G_c(\K)$. Consequently, the compact Feynman diagrams constructed from $G_c(\K)$ collapse onto those of an effective cluster problem embedded in a host which accounts for the fluctuations arising from the hopping of electrons between the cluster and the rest of the system. The compact cluster quantities are then used to calculate the corresponding lattice quantities.

The pairing interaction is given by the irreducible part of the particle-particle vertex 
\begin{equation}
	\label{eq:5} \Gamma^{pp}(K;K')\equiv \Gamma^{pp}(K,-K; K',-K') 
\end{equation}
with $K=(\K,\omega_n)$. One can also use the DCA to calculate the spin susceptibility $\chi(\Q,\omega_n)$ \cite{jarrell:dca3,maier:rev}. We then introduce a d-wave coupling strength \cite{maier:07}
\begin{equation}
	\label{eq:6} -\frac{\frac{1}{2}\langle g(\K) \Gamma^{pp}_{\rm even}(\K,\pi T; \K',\pi T)g(\K')\rangle_{\K\K'}}{\langle g^2(\K)\rangle_{\K}} 
\end{equation}
with the even frequency, even momentum part of the irreducible particle-particle vertex, 
\begin{eqnarray}
	\label{eq:11} \Gamma^{pp}_{\rm even}(\K,\pi T;\K',\pi T) &=& \frac{1}{4} \left( \right. \Gamma^{pp}(\K,\pi T; \K',\pi T)\nonumber\\
	&+&\Gamma^{pp}(\K,\pi T, -\K', \pi T)\nonumber\\
	&+&\Gamma^{pp}(\K,\pi T, \K', -\pi T)\nonumber\\
	&+&\left. \Gamma^{pp}(\K,\pi T, -\K', -\pi T)\right) 
\end{eqnarray}
and $g(\K)=(\cos K_x - \cos K_y)$. 
The leading low temperature eigenvalue of the particle-particle Bethe-Salpeter equation is then calculated from 
\begin{eqnarray}
 -\frac{T}{N_c}\ \sum_{K^\prime} \Gamma^{\rm pp}_{\rm even} \left(K,
 -K; K^\prime, -K^\prime\right)\, {{\bar{\chi}}_0^{\rm pp}}(K') \,
 \phi_\alpha (K^\prime) = &&\nonumber\\ && \hspace{-2cm}\lambda_\alpha
 \phi_\alpha (K)\,. \label{eq:7}
\end{eqnarray}
It is found to correspond to an eigenfunction with d-wave symmetry \cite{maier:pairmech,maier:pairint}. Here we have coarse-grained the Green's function legs, ${{\bar{\chi}}_0^{\rm pp}}(K') = \frac{N_c}{N}\sum_{\tk^{\prime}} G_\uparrow (\K^\prime+\tk^\prime) \, G_\downarrow (-\K^\prime-\tk^\prime)$, according to the DCA assumption. Troughout this paper, we show results calculated on a 2$\times$2 cluster for a near-neighbor hopping $t=1$ and a Hubbard Coulomb interaction $U=8$.

\section{Spin susceptibility representation}\label{sec:spinres}

In Ref.~\onlinecite{maier:07}, we introduced an effective coupling strength $\Ub(T)$ by requiring that the d-wave coupling strength given by Eq.~\eqref{eq:6} be the same at a given temperature when $\Gamma^{pp}(K;K')$ is replaced by the approximate interaction given by Eq.~\eqref{eq:ppRPA}. 
\begin{figure}
	[htbp] \centering 
	\includegraphics[width=3.5in]{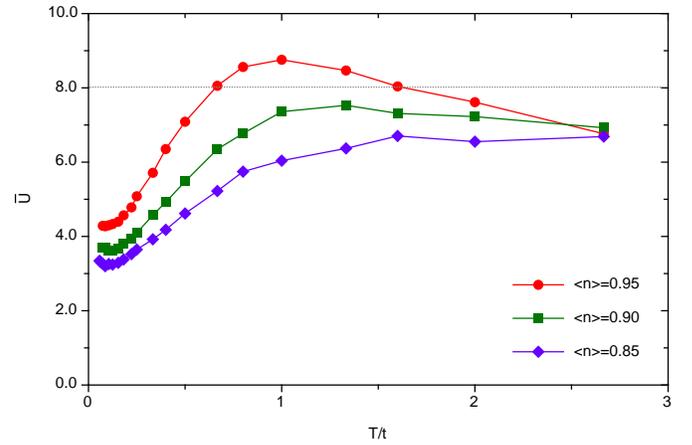} 
	\caption{(Color online) The coupling strength $\Ub(T)$ versus temperature obtained from fitting the ``exact'' DCA QMC pairing interaction Eq.~\eqref{eq:11} for $U=8$ and for different values of the site filling $\n$. We work in units where $t=1$.} 
	\label{fig:Ubarn} 
\end{figure}

In Fig.~\ref{fig:Ubarn} we show the results for $\Ub(T)$ for three different fillings. For temperatures $T<1$, $\Ub(T)$ decreases with temperature for all fillings. One also sees that $\Ub$ decreases with increasing doping for all temperatures. This result is consistent with earlier quantum Monte Carlo calculations that found that the electron-spin fluctuation vertex decreased with decreasing temperature and increasing doping \cite{huang:06}. 

\begin{figure}
	[htbp] \centering 
	\includegraphics[width=3.5in]{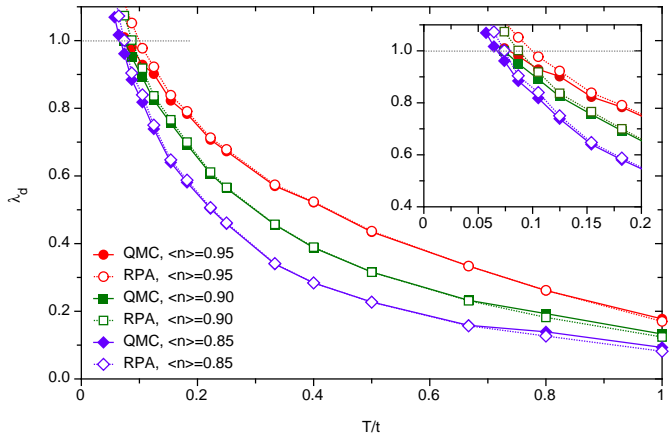} 
	\caption{(Color online) The d-wave eigenvalue $\lambda_d$ versus temperature obtained from the RPA form [Eq.~\eqref{eq:ppRPA}] (open symbols) and from the ``exact'' DCA QMC interaction (solid symbols) for different values of the site filling $\n$.} 
	\label{fig:lambdad} 
\end{figure}

Using these estimates of $\Ub(T)$ one can then explore how well $\frac{3}{2} \Ub^2 \chi(K-K')$ represents $\Gamma^{pp}(K;K')$ by comparing the d-wave eigenvalues. The curves with solid symbols in Fig.~\ref{fig:lambdad} show the d-wave eigenvalue versus $T$ obtained from Eq.~\eqref{eq:7} with the ``exact'' DCA QMC interaction $\Gamma^{pp}$. The curves with open symbols show the d-wave eigenvalue obtained from Eq.~\eqref{eq:7} when $\Gamma^{pp}$ is replaced by $\frac{3}{2}\Ub^2 \chi(K-K')$. In these calculations we have used DCA QMC results for $\chi(K-K')$ as well as the single-particle propagator $G(k)$ that appears in Eq.~\eqref{eq:7}. 
With $\Ub(T)$ determined from fitting $\Gamma^{pp}$, we find for all fillings that the temperature dependence and size of the d-wave eigenvalue $\lambda_d$ are reasonably accounted for by the simple form of the interaction given in Eq.~\eqref{eq:ppRPA}. Since the DCA QMC calculations have been carried out using an $N_c=4$ site cluster, the phase fluctuations are suppressed and the temperature $T_{c0}$ at which the d-wave eigenvalue equals one represents a mean-field transition temperature. In Table~\ref{tab:1} we list the ``exact'' $T_{c0}$ value obtained when $\Gamma^{pp}$ is used to determine $\lambda_d(T)$. The temperature $T_{c0}^{(1)}$ at which $\lambda_d=1$ when $\Gamma^{pp}$ is replaced by $3/2\Ub^2(T)\chi(K-K')$ with $\Ub(T)$ determined from Eq.~\eqref{eq:6} is also listed in Table~\ref{tab:1}. One sees that this approximation over-estimates the mean-field transition temperature by of order 10-30\%, depending upon the doping. Reasons for the disagreement at small doping could include the presence of the pseudogap in the spin excitations and the assumption of a frequency and $\K$ independent coupling strength $\Ub$.

\begin{table}[htpb]
  \centering
  \begin{tabular}{c|llll}
	  $\n$ & $T_{c0}$ & $T_{c0}^{(1)}$ & $T_{c0}^{(2)}$ & $T_{c0}^{(3)}$ \\\hline
	  0.95 & 0.080 & 0.100 (25\%) & 0.108 (35\%) & 0.105 (31\%) \\
	  0.90 & 0.074 & 0.087 (18\%) & 0.084 (14\%) & 0.081 (9\%) \\
	  0.85 & 0.067 & 0.074 (10\%) & 0.064 (4\%) & 0.058 (13\%) 
  \end{tabular}
  \caption{The superconducting mean-field transition temperature obtained as the temperature where the d-wave eigenvalue $\lambda_d=1$ for different values of the site filling $\n$ and different approximations. $T_{c0}$: ``Exact'' result obtained when the DCA QMC result for $\Gamma^{pp}$ is used; $T_{c0}^{(1)}$: result obtained when $\Gamma^{pp}$ is replaced by the RPA form [Eq.~\eqref{eq:ppRPA}] and $\Ub$ is determined from fitting the irreducible particle-particle vertex; $T_{c0}^{(2)}$: result otained when $\Ub$ is determined from fitting the nodel quasiparticle spectral weight using Eq.~\eqref{eq:sigmaRPA}; $T_{c0}^{(3)}$: result obtained when $\Ub$ and $G(K)$ are determined self-consistently by fitting the nodal quasiparticle weight. The numbers in brackets denote the deviation from the ``exact'' result expressed in percent.}
  \label{tab:1}
\end{table}

\section{Single-particle fit of $\Ub(T)$} \label{sec:SFitUb}

In the previous section, we determined the coupling strength $\Ub(T)$ by fitting the pairing interaction. Next we explore how well $\Ub(T)$ can be estimated from the single-particle spectrum by assuming that the self-energy $\Sigma(K)$ is determined by the same form, Eq.~\eqref{eq:ppRPA}, i.e. given by 
\begin{equation}\label{eq:sigmaRPA} 
	-\frac{3}{2}\Ub^2 \sum_{Q} G_c(K-Q) \chi(Q)
\end{equation}
with $K=(\K,i\omega_n)$. As was done for the pairing interaction and the Bethe-Salpeter equation [Eq.~\eqref{eq:7}], we first examine what happens if we use DCA QMC results for the susceptibility $\chi(Q)$ and the single-particle propagator $G_c(K)$. It was shown in Ref.~\onlinecite{macridin:07} that this simple representation of the self-energy provides a useful description of the single-particle spectral weight $A(\k,\omega)$. Within this framework, we propose to estimate the coupling strength $\Ub(T)$ by requiring that the Matsubara quasiparticle weight
\begin{equation}
	\label{eq:Z} Z(\K,T) = \left[1-\frac{\Im m\Sigma(\K,\pi T)}{\pi T}\right]^{-1} 
\end{equation}
for $\K=(\pi/2,\pi/2)$ calculated with the ``exact'' result for $\Sigma(K)$ is the same at a given temperature for the approximate self-energy given by the form in Eq.~\eqref{eq:sigmaRPA}.  

For the 2$\times$2 cluster, the DCA self-energy $\Sigma(K)$ is calculated for the discrete set of momenta $\K=(0,0)$, $(\pi,0)$, $(0,\pi)$ and $(\pi,\pi)$. To obtain the self-energy for $\k=(\pi/2,\pi/2)$ we interpolate $\Sigma(K)$ according to $\Sigma(\k,\omega_n)=\sum_{\bf R}e^{i\k\cdot {\bf R}}\Sigma({\bf R},\omega_n)$ where ${\bf R}$ are the distances in the cluster and $\Sigma({\bf R},\omega_n)$ is the Fourier-transform of $\Sigma(\K,\omega_n)$. This interpolation does not introduce fast Fourier components corresponding to length-scales larger than the cluster, and for the 2$\times$2 cluster is guaranteed to preserve causality. For $\k=(\pi/2,\pi/2)$, one obtains $\Sigma(\pi/2,\pi/2,\omega_n)=\frac{1}{4}\left[ \Sigma(0,0,\omega_n)+2\Sigma(\pi,0,\omega_n)+\Sigma(\pi,\pi,\omega_n)\right]$.

From a phenomenological point of view, it is interesting to see how well the d-wave eigenvalue calculated with the approximate pairing interaction in Eq.~\eqref{eq:ppRPA} and the coupling $\Ub$ determined from the self-energy reproduces the ``exact'' d-wave eigenvalue. To the extent that the 2D Hubbard model gives an appropriate description of the cuprates, this will indicate how well ARPES results can be combined with inelastic neutron scattering results to provide an estimate of the strength of the pairing interaction in the high-T$_c$ cuprates. A similar analysis using Eliashberg equations was applied to the heavy fermion superconductor UPt$_3$ \cite{m_norman_87,w_putikka_89}.

\begin{figure}[htbp]
	\begin{center}
		\includegraphics[width=3.5in]{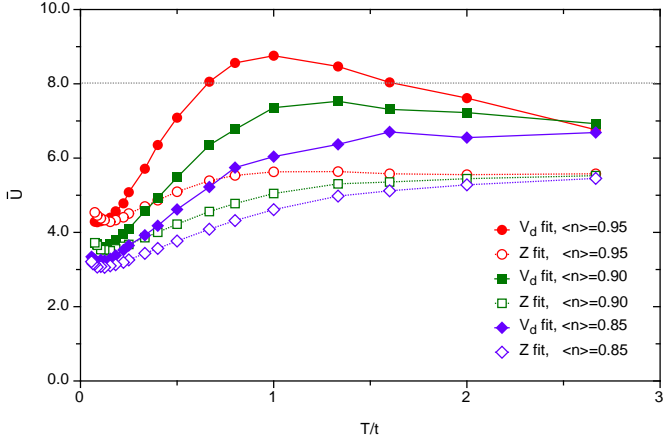}
	\end{center}
	\caption{(Color online) The estimate for the coupling strength $\Ub$ versus temperature determined from fitting the ``exact'' DCA QMC self-energy with the RPA form [Eq.~\eqref{eq:sigmaRPA}] compared to the coupling obtained from fitting the pairing interaction with the RPA form [Eq.~\eqref{eq:ppRPA}] for different site fillings $\n$.}
	\label{fig:UbarComp}
\end{figure}

\begin{figure}[htbp]
	\begin{center}
		\includegraphics[width=3.5in]{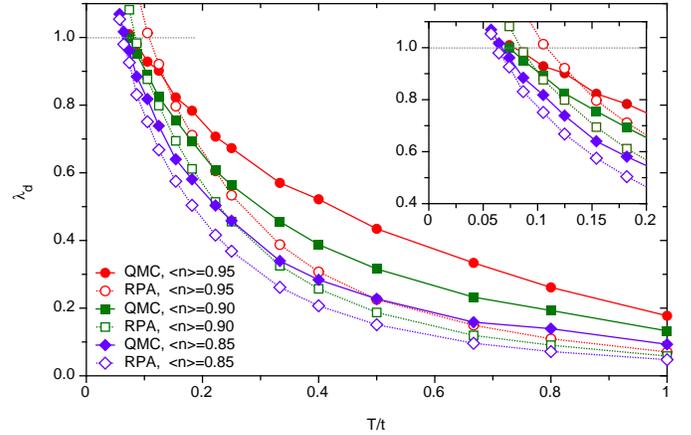}
	\end{center}
	\caption{(Color online) The d-wave eigenvalue $\lambda_d$ versus temperature obtained from the RPA form [Eq.~\eqref{eq:ppRPA}] with $\Ub$ determined from fitting the self-energy with the form [Eq.~\eqref{eq:sigmaRPA}] (open symbols) compared to the eigenvalues obtained from the ``exact'' DCA QMC interaction (solid symbols) for different site fillings $\n$.}
	\label{fig:lambdadUbarfit}
\end{figure}

Fig.~\ref{fig:UbarComp} shows a comparison of the coupling $\Ub$ obtained from fitting the self-energy (open symbols) and the coupling obtained from fitting the pairing interaction as described in Sec.~\ref{sec:spinres} (solid symbols) for various fillings. The corresponding d-wave eigenvalue one obtains by using the approximative form of the pairing interaction, $\frac{3}{2}\Ub^2\chi(K-K')$, in Eq.~\eqref{eq:7} is shown in Fig.~\ref{fig:lambdadUbarfit}. Here, as a comparison, we also plot the ``exact'' results for $\lambda_d$ obtained if the DCA QMC results for the pairing interaction $\Gamma^{pp}$ are used in Eq.~\eqref{eq:7}. In Fig.~\ref{fig:UbarComp} one sees that the single-particle estimate of $\Ub$ is smaller than the that obtained from fitting the pairing interaction for all temperatures and fillings. The result of this is that the corresponding eigenvalue obtained in this approximation can be larger or smaller than the ``exact'' eigenvalue, depending upon the doping and the temperature. Nevertheless, as shown in Table~\ref{tab:1}, the errors in the determination of the mean-field transition temperature are similar in size to the case in which $\Ub$ was determined by a fit that required knowledge of $\Gamma^{pp}$.

\section{Self-consistent determination of $\Ub(T)$ and $G(K)$}\label{sec:selfcFit}

In the previous sections, we used DCA QMC results for the spin susceptibility $\chi(Q)$ and the single-particle Green's function $G(K)$, and estimated the coupling strength $\Ub(T)$ either by fitting the pairing interaction or the nodal quasiparticle weight. In this section, we go one step further and assume a scenario where only limited information is available for the single-particle Green's function $G(K)$. In this case, we use the approximate form [Eq.~\eqref{eq:sigmaRPA}] for the self-energy $\Sigma(K)$ to determine the dressed Green's function from the DCA coarse grained Dyson equation
\begin{eqnarray}
	G^{-1}_c(\K,\omega_n)=\frac{N_c}{N}\sum_\kt \left[ G_0^{-1}(\K+\kt,\omega_n)-\Sigma(\K,\omega_n)\right]^{-1}\,.
	\label{eq:dyson}
\end{eqnarray}
Here, $G_0(k)$ is the non-interacting Green's function, i.e. $G_0(\k,\omega_n)=(i\omega_n-\epsilon_\k)^{-1}$ with $\epsilon_\k=-2t (\cos k_x + \cos k_y)$. Eqs.~\eqref{eq:sigmaRPA} and \eqref{eq:dyson} are iterated until self-consistency is achieved, and the value of $\Ub$ entering Eq.~\eqref{eq:sigmaRPA} is again fixed by requiring that the quasiparticle weight $Z(\K,T)$ obtained with this approximation for $\K=(\pi/2,\pi/2)$ is the same as that obtained from the  ``exact'' DCA QMC self-energy. We find that the estimates one obtains for $\Ub$ using this approach are almost identical to the values obtained in Sec.~\ref{sec:SFitUb} (see Fig.~\ref{fig:UbarComp}) and therefore do not show the results.

\begin{figure}[htpb]
	\begin{center}
		\includegraphics[width=3.5in]{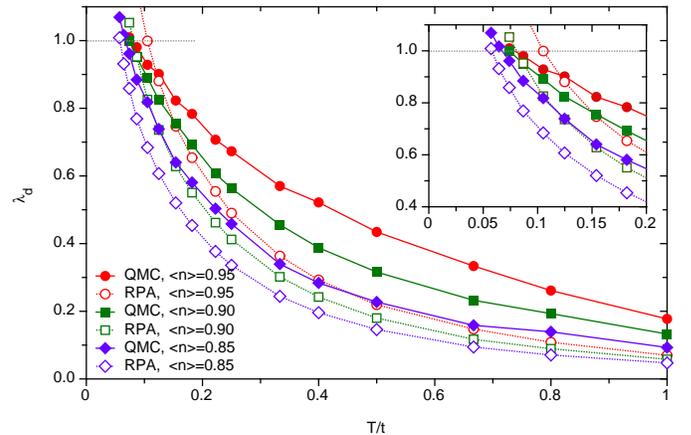}
	\end{center}
	\caption{(Color online) The d-wave eigenvalue $\lambda_d$ versus temperature obtained from the RPA form [Eq.~\eqref{eq:ppRPA}] with $\Ub$ and $G(K)$ determined self-consistently from fitting the self-energy with the approximate form [Eq.~\eqref{eq:sigmaRPA}] (open symbols) compared to the eigenvalues obtained from the ``exact'' DCA QMC interaction (solid symbols) for different site fillings $\n$.}
	\label{fig:lambdadUbarGfit}
\end{figure}

Using these estimates of $\Ub(T)$ and $G(k)$, one can again calculate the d-wave eigenvalue and compare it to the ``exact'' eigenvalue. The curves with open symbols in Fig.~\ref{fig:lambdadUbarGfit} show the result for the d-wave eigenvalue obtained with this approximation and the curves with solid symbols display the ``exact'' result. One sees that the results obtained with this approximation are almost identical to the results obtained in Sec.~\ref{sec:SFitUb} where the ``exact'' DCA QMC Green's function was used. In this case, the temperatures at which $\lambda_d(T)=1$ are listed as $T_{c0}^{(3)}$ in Table~\ref{tab:1} and one sees that they are again within 10\% to 30\% of the exact $T_{c0}$ values. Apparently the additional renormalized one-loop approximation [Eq.~\eqref{eq:sigmaRPA}] of the self-energy that enters the Bethe-Salpether equation [Eq.~\eqref{eq:7}] through the propagator $G(k)$ has only a negligible effect on the d-wave eigenvalue.

We also performed fits of the real frequency spectra with the corresponding one-loop approximation. On small clusters we find it difficult to describe the spectra with this simple approximation. On larger clusters, however, we have previously shown that the spectra are well approximated by the one-loop form for the self-energy \cite{macridin:07}.  Preliminary results on a 16-site cluster indicate that very similar estimates of Ubar are obtained as those presented in this manuscript.

\section{Conclusions}\label{sec:concl}

In conclusion, this work has shown that the pairing interaction in the 2D Hubbard model in the parameter regime appropriate for the cuprates over a range of dopings is well described by the spin susceptibility representation, $\frac{3}{2}\Ub(T)^2\chi(K-K')$. At low temperatures, the coupling strength $\Ub$ decreases with temperature and doping. Close to the superconducting transition temperature, $\Ub$ can be well estimated by assuming that the self-energy is determined by the same interaction, $\frac{3}{2}\Ub(T)^2\chi(Q)$, and requiring that one has the same nodal quasiparticle weight as the ``exact'' DCA QMC result. In practice, one would seek to relate this to the renormalization of the nodal Fermi velocity measured in ARPES studies. Using this approximation to self-consistently determine the single-particle propagator that enters the Bethe-Salpeter equation in the particle-particle channel has only neglibible effects on its d-wave eigenvalue. As a result, estimates of the superconducting mean-field transition temperature using the spin-susceptibility representation of the pairing interaction and the self-energy provide a satisfactory way of measuring the strength of the d-wave pairing interaction.

\acknowledgments This research was enabled by computational resources of the Center for Computational Sciences at Oak Ridge National Laboratory. TAM and DJS acknowledge the Center for Nanophase Materials Sciences, which is sponsored at Oak Ridge National Laboratory by the Division of Scientific User Facilities, U.S. Department of Energy. Work at Cincinnati was supported by the NSF under grant NSF DMR-0312680 and the DOE under grant CMSN DOE DE-FG02-04ER46129.

\bibliography{mybib}

\begin{thebibliography}{21}
\expandafter\ifx\csname natexlab\endcsname\relax\def\natexlab#1{#1}\fi
\expandafter\ifx\csname bibnamefont\endcsname\relax
  \def\bibnamefont#1{#1}\fi
\expandafter\ifx\csname bibfnamefont\endcsname\relax
  \def\bibfnamefont#1{#1}\fi
\expandafter\ifx\csname citenamefont\endcsname\relax
  \def\citenamefont#1{#1}\fi
\expandafter\ifx\csname url\endcsname\relax
  \def\url#1{\texttt{#1}}\fi
\expandafter\ifx\csname urlprefix\endcsname\relax\def\urlprefix{URL }\fi
\providecommand{\bibinfo}[2]{#2}
\providecommand{\eprint}[2][]{\url{#2}}

\bibitem[{\citenamefont{Maier et~al.}(2006{\natexlab{a}})\citenamefont{Maier,
  Jarrell, and Scalapino}}]{maier:pairmech}
\bibinfo{author}{\bibfnamefont{T.}~\bibnamefont{Maier}},
  \bibinfo{author}{\bibfnamefont{M.}~\bibnamefont{Jarrell}}, \bibnamefont{and}
  \bibinfo{author}{\bibfnamefont{D.}~\bibnamefont{Scalapino}},
  \bibinfo{journal}{Phys. Rev. Lett.} \textbf{\bibinfo{volume}{96}},
  \bibinfo{pages}{047005} (\bibinfo{year}{2006}{\natexlab{a}}).

\bibitem[{\citenamefont{Maier et~al.}(2006{\natexlab{b}})\citenamefont{Maier,
  Jarrell, and Scalapino}}]{maier:pairint}
\bibinfo{author}{\bibfnamefont{T.}~\bibnamefont{Maier}},
  \bibinfo{author}{\bibfnamefont{M.}~\bibnamefont{Jarrell}}, \bibnamefont{and}
  \bibinfo{author}{\bibfnamefont{D.}~\bibnamefont{Scalapino}},
  \bibinfo{journal}{Phys. Rev. B} \textbf{\bibinfo{volume}{74}},
  \bibinfo{pages}{094513} (\bibinfo{year}{2006}{\natexlab{b}}).

\bibitem[{\citenamefont{Maier et~al.}(2007)\citenamefont{Maier, Jarrell, and
  Scalapino}}]{maier:07}
\bibinfo{author}{\bibfnamefont{T.}~\bibnamefont{Maier}},
  \bibinfo{author}{\bibfnamefont{M.}~\bibnamefont{Jarrell}}, \bibnamefont{and}
  \bibinfo{author}{\bibfnamefont{D.}~\bibnamefont{Scalapino}},
  \bibinfo{journal}{Phys. Rev. B} \textbf{\bibinfo{volume}{75}},
  \bibinfo{pages}{134519} (\bibinfo{year}{2007}).

\bibitem[{\citenamefont{Berk and Schrieffer}(1966)}]{berk:66}
\bibinfo{author}{\bibfnamefont{N.}~\bibnamefont{Berk}} \bibnamefont{and}
  \bibinfo{author}{\bibfnamefont{J.}~\bibnamefont{Schrieffer}},
  \bibinfo{journal}{Phys. Rev. Lett.} \textbf{\bibinfo{volume}{17}},
  \bibinfo{pages}{433} (\bibinfo{year}{1966}).

\bibitem[{\citenamefont{Scalapino}(1995)}]{scalapino:95}
\bibinfo{author}{\bibfnamefont{D.}~\bibnamefont{Scalapino}},
  \bibinfo{journal}{Phys. Rep.} \textbf{\bibinfo{volume}{250}},
  \bibinfo{pages}{329} (\bibinfo{year}{1995}).

\bibitem[{\citenamefont{Moriya and Ueda}(2000)}]{moriya:00}
\bibinfo{author}{\bibfnamefont{T.}~\bibnamefont{Moriya}} \bibnamefont{and}
  \bibinfo{author}{\bibfnamefont{K.}~\bibnamefont{Ueda}},
  \bibinfo{journal}{Adv. in Phys.} \textbf{\bibinfo{volume}{49}},
  \bibinfo{pages}{555} (\bibinfo{year}{2000}).

\bibitem[{\citenamefont{Chubokov et~al.}(2003)\citenamefont{Chubokov, Pines,
  and Schmalian}}]{chubokov:03}
\bibinfo{author}{\bibfnamefont{A.}~\bibnamefont{Chubokov}},
  \bibinfo{author}{\bibfnamefont{D.}~\bibnamefont{Pines}}, \bibnamefont{and}
  \bibinfo{author}{\bibfnamefont{J.}~\bibnamefont{Schmalian}},
  \emph{\bibinfo{title}{The Physics of Superconductors: Conventional and
  High-T$_c$ Superconductors}} (\bibinfo{publisher}{Springer-Verlag},
  \bibinfo{year}{2003}), chap.~\bibinfo{chapter}{7}.

\bibitem[{\citenamefont{Hettler et~al.}(1998)\citenamefont{Hettler,
  Tahvildar-Zadeh, Jarrell, Pruschke, and Krishnamurthy}}]{hettler:dca1}
\bibinfo{author}{\bibfnamefont{M.~H.} \bibnamefont{Hettler}},
  \bibinfo{author}{\bibfnamefont{A.~N.} \bibnamefont{Tahvildar-Zadeh}},
  \bibinfo{author}{\bibfnamefont{M.}~\bibnamefont{Jarrell}},
  \bibinfo{author}{\bibfnamefont{T.}~\bibnamefont{Pruschke}}, \bibnamefont{and}
  \bibinfo{author}{\bibfnamefont{H.~R.} \bibnamefont{Krishnamurthy}},
  \bibinfo{journal}{Phys. Rev. B} \textbf{\bibinfo{volume}{58}},
  \bibinfo{pages}{R7475} (\bibinfo{year}{1998}).

\bibitem[{\citenamefont{Hettler et~al.}(2000)\citenamefont{Hettler, Mukherjee,
  Jarrell, and Krishnamurthy}}]{hettler:dca2}
\bibinfo{author}{\bibfnamefont{M.~H.} \bibnamefont{Hettler}},
  \bibinfo{author}{\bibfnamefont{M.}~\bibnamefont{Mukherjee}},
  \bibinfo{author}{\bibfnamefont{M.}~\bibnamefont{Jarrell}}, \bibnamefont{and}
  \bibinfo{author}{\bibfnamefont{H.~R.} \bibnamefont{Krishnamurthy}},
  \bibinfo{journal}{Phys. Rev. B} \textbf{\bibinfo{volume}{61}},
  \bibinfo{pages}{12739} (\bibinfo{year}{2000}).

\bibitem[{\citenamefont{Jarrell
  et~al.}(2001{\natexlab{a}})\citenamefont{Jarrell, Maier, Huscroft, and
  Moukouri}}]{jarrell:dca3}
\bibinfo{author}{\bibfnamefont{M.}~\bibnamefont{Jarrell}},
  \bibinfo{author}{\bibfnamefont{T.}~\bibnamefont{Maier}},
  \bibinfo{author}{\bibfnamefont{C.}~\bibnamefont{Huscroft}}, \bibnamefont{and}
  \bibinfo{author}{\bibfnamefont{S.}~\bibnamefont{Moukouri}},
  \bibinfo{journal}{Phys. Rev. B} \textbf{\bibinfo{volume}{64}},
  \bibinfo{pages}{195130} (\bibinfo{year}{2001}{\natexlab{a}}).

\bibitem[{\citenamefont{Maier et~al.}(2005{\natexlab{a}})\citenamefont{Maier,
  Jarrell, Pruschke, and Hettler}}]{maier:rev}
\bibinfo{author}{\bibfnamefont{T.}~\bibnamefont{Maier}},
  \bibinfo{author}{\bibfnamefont{M.}~\bibnamefont{Jarrell}},
  \bibinfo{author}{\bibfnamefont{T.}~\bibnamefont{Pruschke}}, \bibnamefont{and}
  \bibinfo{author}{\bibfnamefont{M.}~\bibnamefont{Hettler}},
  \bibinfo{journal}{Rev. Mod. Phys.} \textbf{\bibinfo{volume}{77}},
  \bibinfo{pages}{1027} (\bibinfo{year}{2005}{\natexlab{a}}).

\bibitem[{\citenamefont{Maier et~al.}(2000)\citenamefont{Maier, Jarrell,
  Pruschke, and Keller}}]{maier:dca1}
\bibinfo{author}{\bibfnamefont{T.}~\bibnamefont{Maier}},
  \bibinfo{author}{\bibfnamefont{M.}~\bibnamefont{Jarrell}},
  \bibinfo{author}{\bibfnamefont{T.}~\bibnamefont{Pruschke}}, \bibnamefont{and}
  \bibinfo{author}{\bibfnamefont{J.}~\bibnamefont{Keller}},
  \bibinfo{journal}{Eur. Phys. J B} \textbf{\bibinfo{volume}{13}},
  \bibinfo{pages}{613} (\bibinfo{year}{2000}).

\bibitem[{\citenamefont{Jarrell
  et~al.}(2001{\natexlab{b}})\citenamefont{Jarrell, Maier, Hettler, and
  Tahvildarzadeh}}]{jarrell:dca2}
\bibinfo{author}{\bibfnamefont{M.}~\bibnamefont{Jarrell}},
  \bibinfo{author}{\bibfnamefont{T.}~\bibnamefont{Maier}},
  \bibinfo{author}{\bibfnamefont{M.~H.} \bibnamefont{Hettler}},
  \bibnamefont{and} \bibinfo{author}{\bibfnamefont{A.~N.}
  \bibnamefont{Tahvildarzadeh}}, \bibinfo{journal}{Europhys. Lett.}
  \textbf{\bibinfo{volume}{56}}, \bibinfo{pages}{563}
  (\bibinfo{year}{2001}{\natexlab{b}}).

\bibitem[{\citenamefont{Macridin et~al.}(2006)\citenamefont{Macridin, Jarrell,
  Maier, and Kent}}]{macridin:06}
\bibinfo{author}{\bibfnamefont{A.}~\bibnamefont{Macridin}},
  \bibinfo{author}{\bibfnamefont{M.}~\bibnamefont{Jarrell}},
  \bibinfo{author}{\bibfnamefont{T.}~\bibnamefont{Maier}}, \bibnamefont{and}
  \bibinfo{author}{\bibfnamefont{P.}~\bibnamefont{Kent}},
  \bibinfo{journal}{Phys. Rev. Lett.} \textbf{\bibinfo{volume}{97}},
  \bibinfo{pages}{036401} (\bibinfo{year}{2006}).

\bibitem[{\citenamefont{Kyung et~al.}(2006)\citenamefont{Kyung, Kancharla,
  S\'en\'echal, and Tremblay}}]{kyung:06}
\bibinfo{author}{\bibfnamefont{B.}~\bibnamefont{Kyung}},
  \bibinfo{author}{\bibfnamefont{S.}~\bibnamefont{Kancharla}},
  \bibinfo{author}{\bibfnamefont{D.}~\bibnamefont{S\'en\'echal}},
  \bibnamefont{and} \bibinfo{author}{\bibfnamefont{A.-M.}
  \bibnamefont{Tremblay}}, \bibinfo{journal}{Phys. Rev. B}
  \textbf{\bibinfo{volume}{73}}, \bibinfo{pages}{165114}
  (\bibinfo{year}{2006}).

\bibitem[{\citenamefont{S\'en\'echal and Tremblay}(2004)}]{senechal:04a}
\bibinfo{author}{\bibfnamefont{D.}~\bibnamefont{S\'en\'echal}}
  \bibnamefont{and} \bibinfo{author}{\bibfnamefont{A.-M.}
  \bibnamefont{Tremblay}}, \bibinfo{journal}{Phys. Rev. Lett.}
  \textbf{\bibinfo{volume}{92}}, \bibinfo{pages}{126401}
  (\bibinfo{year}{2004}).

\bibitem[{\citenamefont{Maier et~al.}(2005{\natexlab{b}})\citenamefont{Maier,
  Jarrell, Schulthess, Kent, and White}}]{maier:schm}
\bibinfo{author}{\bibfnamefont{T.}~\bibnamefont{Maier}},
  \bibinfo{author}{\bibfnamefont{M.}~\bibnamefont{Jarrell}},
  \bibinfo{author}{\bibfnamefont{T.}~\bibnamefont{Schulthess}},
  \bibinfo{author}{\bibfnamefont{P.}~\bibnamefont{Kent}}, \bibnamefont{and}
  \bibinfo{author}{\bibfnamefont{J.}~\bibnamefont{White}},
  \bibinfo{journal}{Phys. Rev. Lett.} \textbf{\bibinfo{volume}{95}},
  \bibinfo{pages}{237001} (\bibinfo{year}{2005}{\natexlab{b}}).

\bibitem[{\citenamefont{Huang et~al.}(2006)\citenamefont{Huang, Hanke,
  Arrigoni, and Chubukov}}]{huang:06}
\bibinfo{author}{\bibfnamefont{Z.}~\bibnamefont{Huang}},
  \bibinfo{author}{\bibfnamefont{W.}~\bibnamefont{Hanke}},
  \bibinfo{author}{\bibfnamefont{E.}~\bibnamefont{Arrigoni}}, \bibnamefont{and}
  \bibinfo{author}{\bibfnamefont{A.}~\bibnamefont{Chubukov}},
  \bibinfo{journal}{Phys. Rev. B} \textbf{\bibinfo{volume}{74}},
  \bibinfo{pages}{184508} (\bibinfo{year}{2006}).

\bibitem[{\citenamefont{Macridin et~al.}(2007)\citenamefont{Macridin, Jarrell,
  Maier, and Scalapino}}]{macridin:07}
\bibinfo{author}{\bibfnamefont{A.}~\bibnamefont{Macridin}},
  \bibinfo{author}{\bibfnamefont{M.}~\bibnamefont{Jarrell}},
  \bibinfo{author}{\bibfnamefont{T.}~\bibnamefont{Maier}}, \bibnamefont{and}
  \bibinfo{author}{\bibfnamefont{D.}~\bibnamefont{Scalapino}},
  \bibinfo{journal}{preprint, cond-mat/0701429}  (\bibinfo{year}{2007}).

\bibitem[{\citenamefont{Norman}(1987)}]{m_norman_87}
\bibinfo{author}{\bibfnamefont{M.}~\bibnamefont{Norman}},
  \bibinfo{journal}{Phys. Rev. Lett.} \textbf{\bibinfo{volume}{59}},
  \bibinfo{pages}{232} (\bibinfo{year}{1987}).

\bibitem[{\citenamefont{Putikka and Joynt}(1989)}]{w_putikka_89}
\bibinfo{author}{\bibfnamefont{W.~O.} \bibnamefont{Putikka}} \bibnamefont{and}
  \bibinfo{author}{\bibfnamefont{R.}~\bibnamefont{Joynt}},
  \bibinfo{journal}{Phys. Rev. B} \textbf{\bibinfo{volume}{39}},
  \bibinfo{pages}{701 } (\bibinfo{year}{1989}).

\end{thebibliography}

\end{document}